\documentclass{article}

\pdfoutput=1
\usepackage{PRIMEarxiv}

\usepackage[utf8]{inputenc} 
\usepackage[T1]{fontenc}    
\usepackage{hyperref}       
\usepackage{url}            
\usepackage{booktabs}       
\usepackage{amsfonts}       
\usepackage{nicefrac}       
\usepackage{microtype}      
\usepackage{lipsum}
\usepackage{fancyhdr}       
\usepackage{graphicx}       
\graphicspath{{media/}}     
\usepackage{natbib}
\setcitestyle{authoryear} 
\usepackage[figuresright]{rotating}
\usepackage{amsmath}
\usepackage{amsfonts}

\usepackage[T1]{fontenc}
\usepackage{newpxmath}
\usepackage{hyperref}
\hypersetup{
    colorlinks,
    linkcolor={blue},
    citecolor={blue},
    urlcolor={black}
}

\usepackage{lscape}
\usepackage{longtable}
\usepackage{multirow}
\usepackage{threeparttable}

\title{Capturing Cumulative Disease Burden in Chronic Kidney Disease Outcome Trials: Area Under the Curve and Restricted Mean Time in Favor of Treatment Beyond Conventional Time-to-First Analysis}

 \author{
   Jiren Sun \\
   Global Statistical Science \\
   Eli Lilly and Company \\
   Indianapolis, IN, 46285 \\
   \texttt{jiren.sun@lilly.com} \\
    \AND
    Tuo Wang \\
   Global Statistical Science \\
   Eli Lilly and Company \\
   Indianapolis, IN, 46285 \\
     \texttt{tuo.wang@lilly.com} \\
       \And
    Yu Du \\
   Global Statistical Science \\
   Eli Lilly and Company \\
   Indianapolis, IN, 46285 \\
    \texttt{du\_yu@lilly.com} \\
 }

\begin{document}
\maketitle

\begin{abstract}
\textbf{Background:} Chronic kidney disease (CKD) affects millions worldwide and progresses irreversibly through stages culminating in end-stage renal disease (ESRD) and death. Outcome trials in CKD traditionally employ time-to-first-event analyses using Cox proportional hazards models. However, this approach has fundamental limitations for progressive diseases: it assigns equal weight to each composite endpoint component despite clear clinical hierarchy---an eGFR decline threshold (whether 40\%, 50\%, or 57\%, each validated as surrogates for kidney failure) receives the same weight as ESRD or death in the analysis, and it captures only the first occurrence while ignoring subsequent progression. Given CKD's gradual evolution over years, comprehensive treatment evaluation requires quantifying cumulative disease burden---integrating both event severity and time spent in each disease state.

\textbf{Methods:} We propose two complementary approaches to better characterize treatment benefits by incorporating event severity and state occupancy: area under the curve (AUC) and restricted mean time in favor of treatment (RMT-IF). The AUC method assigns ordinal severity scores to disease states and calculates the area under the mean cumulative score curve, quantifying total event-free time lost. Treatment effects are expressed as AUC ratios or differences. The RMT-IF extends restricted mean survival time to multistate processes, measuring average time patients in the treatment arm spend in more favorable states versus the comparator. We demonstrate both methods using REWIND trial data comparing dulaglutide with placebo in 4,133 adults with type 2 diabetes and CKD (baseline eGFR $<$ 60 mL/min/1.73 m$^2$ or UACR $>$ 30 mg/g). We analyzed five hierarchical renal outcomes: 40\%, 50\%, and 57\% eGFR declines (as commonly applied in CKD outcome trials indicating meaningful disease progression), ESRD, and all-cause mortality.

\textbf{Results:} Among 4,133 participants (2,040 dulaglutide; 2,093 placebo), conventional analysis yielded a composite hazard ratio of 0.899 (95\% CI: 0.806--1.004; $p = 0.06$). Over six years, the AUC ratio was 0.857 (95\% CI: 0.760--0.966; $p = 0.012$), indicating 14.3\% reduction in total event-time that would have been lost due to CKD disease progression. The RMT-IF was 0.116 years ($p = 0.018$), representing 42 additional days in favorable health states, including 29 days of extended survival. Sensitivity analyses demonstrated remarkable stability of AUC estimates (0.851--0.857) across different eGFR threshold combinations, contrasting with substantial variability in time-to-first hazard ratios (0.861--0.899).

\textbf{Conclusion:} The AUC and RMT-IF approaches overcame critical limitations of conventional time-to-first analyses by integrating disease severity with state duration, providing clinically interpretable measures of cumulative burden. These methods better capture CKD's progressive nature where treatment benefits extend beyond first-event delay to overall disease trajectory modification. By discriminating between events of differing clinical importance and quantifying the complete disease course, these estimands offer alternative assessment frameworks for kidney-protective therapies, potentially improving efficiency and interpretability of future CKD outcome trials.

\end{abstract}

\keywords{Area under the curve \and Chronic kidney disease \and Composite endpoints \and Cumulative disease burden \and eGFR decline thresholds \and Multistate models \and Restricted mean time in favor of treatment \and Clinical trial methodology}

\section{Introduction}

Chronic kidney disease (CKD) represents a major global health challenge affecting approximately 10\% of the world's population, with prevalence continuing to rise due to aging populations and increasing rates of diabetes and obesity \citep{jha2013chronic,webster2017chronic}. The disease follows a predictable trajectory of progressive and irreversible kidney function decline, ultimately culminating in end-stage renal disease (ESRD) requiring dialysis or transplantation, with substantially elevated mortality risk throughout this progression \citep{go2004chronic}. This inexorable course places enormous burdens on patients, families, and healthcare systems worldwide.

Clinical trials evaluating kidney-protective therapies have traditionally incorporated composite renal endpoints combining ESRD with surrogate markers of disease progression. The established surrogate of doubling serum creatinine---corresponding to approximately 57\% reduction in estimated glomerular filtration rate (eGFR)---has long served as a validated endpoint predictive of progression to kidney failure \citep{levey2014gfr}. However, these severe outcomes typically manifest only in late-stage CKD, necessitating extended follow-up periods and large sample sizes to demonstrate treatment effects.

Recent collaborative efforts between regulatory agencies and the nephrology community have validated smaller eGFR decline thresholds as reliable surrogate endpoints. A landmark workshop convened by the National Kidney Foundation and FDA comprehensively evaluated evidence supporting 40\% and 50\% eGFR declines as valid indicators of CKD progression \citep{inker2019gfr}. These thresholds are now commonly applied across contemporary CKD outcome trials, each representing meaningful disease progression. The CREDENCE trial utilized a 57\% eGFR decline in its primary composite endpoint \citep{perkovic2019canagliflozin}, while DAPA-CKD incorporated a 50\% decline \citep{heerspink2020dapagliflozin}, and more recent trials including EMPA-KIDNEY have adopted the 40\% threshold \citep{empa2023empagliflozin}. Importantly, these thresholds exhibit a hierarchical relationship---patients experiencing a 57\% decline must have necessarily progressed through 40\% and 50\% declines, reflecting the continuous and progressive nature of kidney function loss. Meta-analyses across multiple trials have confirmed that treatment effects on these smaller decline thresholds closely parallel effects on traditional ``hard'' endpoints while occurring more frequently, thereby improving trial feasibility \citep{heerspink2023effects}. Evidence suggests that adopting a 40\% decline threshold requires approximately half the sample size of a 57\% decline while maintaining comparable treatment effect estimates \citep{cherney2021kidney}.

Despite these advances in endpoint selection, the analytical framework for evaluating treatment efficacy remains predominantly anchored to time-to-first-event analyses using Cox proportional hazards models. While this approach has proven valuable across many disease areas, it presents specific limitations when applied to chronic progressive conditions like CKD. First, the method treats each component event equally despite their clear clinical hierarchy---a 40\% eGFR decline represents a fundamentally different clinical state than ESRD or death, yet both contribute equally to the composite hazard ratio. Second, by focusing exclusively on first events, the analysis ignores subsequent disease progression that significantly impacts overall disease burden and patient outcomes.

Consider two illustrative patient scenarios: Patient A experiences a 40\% eGFR decline at year one but subsequently stabilizes without further progression. Patient B experiences the same 40\% decline at year two but continues deteriorating through 50\% and 57\% declines to ESRD and death. Under conventional time-to-first analysis, Patient B appears to have the better outcome due to later first-event timing, despite experiencing far worse overall disease trajectory. This fundamental disconnect between analytical approach and clinical reality undermines our ability to fully capture treatment benefits in progressive diseases.

The concept of cumulative disease burden offers a more comprehensive framework for evaluating interventions in CKD. Recent methodological advances in cardiovascular trials have demonstrated the value of integrating multiple events over time to quantify total disease impact \citep{claggett2018quantifying,gregson2023recurrent}. The area under the curve (AUC) approach, successfully applied in heart failure trials, provides interpretable measures of cumulative event-free time lost \citep{claggett2018quantifying}. Similarly, the restricted mean time in favor of treatment (RMT-IF) extends familiar restricted mean survival time (RMST) concepts to multistate disease processes \citep{mao2023restricted}.

In this manuscript, we present and compare two complementary analytical approaches---AUC and RMT-IF---that address limitations of conventional time-to-first analyses by incorporating both event severity and duration in each disease state. Using data from the REWIND trial's CKD risk population, we demonstrate how these methods provide more comprehensive assessment of treatment effects, better reflecting the chronic progressive nature of kidney disease and the full scope of therapeutic benefit. The remainder of this manuscript is organized as follows: Section \ref{sec:method} details the mathematical framework and estimation procedures for both methods; Section \ref{sec:results} presents results from the REWIND trial analysis including sensitivity analyses; and Section \ref{sec:discussion} discusses clinical implications, methodological advantages, future trial design considerations, limitations, and concluding perspectives on the role of these methods in advancing CKD therapeutic evaluation.

\section{Method}\label{sec:method}
\subsection{Mathematical Framework and Notations}
We define the analytical framework over a prespecified time interval $[0, \tau]$ for comparing treatment arms. Let $Y^{(a)}(t)$ represent the multistate outcome process for treatment arm $a \in \{1, 0\}$, where 1 denotes active treatment and 0 denotes placebo. The state space comprises $\{0, 1, \ldots, K, K+1\}$, where state 0 represents the baseline state without events (censoring), states 1 through $K$ represent increasingly severe renal outcomes, and state $K+1$ represents death.

For our CKD application with five hierarchical outcomes (40\%, 50\%, 57\% eGFR decline, ESRD, and death), we have $K = 4$. The multistate process exhibits progressive characteristics, meaning $Y^{(a)}(t) \leq Y^{(a)}(s)$ for all $0 \leq t \leq s \leq \tau$, reflecting CKD's irreversible nature. Transition times into state $k$ or more severe states are defined as $T_k^{(a)} = \inf\{t : Y^{(a)}(t) \geq k\}$, for $k = 1, \ldots, K+1$. By construction, $T^{(a)}_1 \le \cdots \le T^{(a)}_{K+1}$, with equality possible due to ``state skipping''. For instance, if a subject is assessed periodically and is first observed to have a $57\%$ decline in eGFR at time $t$ without prior records of $40\%$ or $50\%$ eGFR declines, then $T^{(a)}_1 = T^{(a)}_2 = T^{(a)}_3 = t$. In this case, the times of $40\%$ and $50\%$ eGFR decline are left-censored at $t$.

Given the realities of clinical observation, we account for censoring through $C^{(a)}$, assumed independent of the outcome process. The observed data comprise censored transition times $\{X_k^{(a)}, \delta_k^{(a)}\}$, where $X_k^{(a)} = T_k^{(a)} \wedge C^{(a)}$ and $\delta_k^{(a)} = I\{T_k^{(a)} \leq C^{(a)}\}$, with $I(\cdot)$ denoting the indicator function, and $a \wedge b = \min(a, b)$.

\subsection{Area Under the Curve (AUC) Methodology}
The AUC approach quantifies cumulative disease burden through explicit severity scoring. We assign ordinal scores 1 through 5 to the five renal events (40\%, 50\%, 57\% eGFR decline, ESRD, and death), reflecting their increasing clinical severity. These scores define a counting process $N^{(a)}(t)$ representing the patient's current disease state:
\begin{equation*}
N^{(a)}(t) = (K + 1) - \sum_{k=1}^{K+1} I\{T_k^{(a)} > t\}
\end{equation*}

This formulation ensures $N^{(a)}(t)$ increases monotonically as patients progress through disease states, capturing the irreversible nature of CKD progression. The area under this trajectory, $A^{(a)}(t)$, quantifies cumulative burden:
\begin{equation*}
A^{(a)}(t) = (K + 1)t - \sum_{k=1}^{K+1} A_k^{(a)}(t)
\end{equation*}
where $A_k^{(a)}(t) = \min\{T_k^{(a)}, t\}$. At the population level, $E\{A_k^{(a)}(t)\}$ corresponds to the RMST of $T^{(a)}_k$ up to time $t$; that is, the expected event-free survival time for the $k$th transition, truncated at $t$. Accordingly, $E\{A^{(a)}(t)\}$ can be interpreted as the total event-free time lost due to disease progression, and thus provides a quantitative summary of cumulative disease burden.

Treatment effects are summarized through two complementary measures:
\begin{itemize}
\item \textbf{AUC Ratio:} $R(\tau) = E\{A^{(1)}(\tau)\}/E\{A^{(0)}(\tau)\}$, providing a relative measure of burden reduction
\item \textbf{AUC Difference:} $D(\tau) = E\{A^{(1)}(\tau)\} - E\{A^{(0)}(\tau)\}$, quantifying absolute burden reduction in event-years
\end{itemize}

Negative values of $\log R(\tau)$ or $D(\tau)$ indicate treatment benefit through reduced cumulative disease burden as reflected in the total event-time that would have been lost due to CKD disease progression.

\subsubsection{Estimation and Statistical Inference}
Given observed data from $n_a$ participants in arm $a$, we estimate population quantities using nonparametric methods. The expected cumulative score $E\{N^{(a)}(t)\}$ is estimated as:
\begin{equation*}
\hat{E}\{N^{(a)}(t)\} = (K + 1) - \sum_{k=1}^{K+1} \hat{S}_k^{(a)}(t)
\end{equation*}
where $\hat{S}_k^{(a)}(\cdot)$ denotes the Kaplan-Meier (KM) estimator for the survival function of $T_k^{(a)}$. The AUC estimate follows as:
\begin{equation*}
\hat{E}\{A^{(a)}(t)\} = (K + 1)t - \sum_{k=1}^{K+1} \hat{E}\{A_k^{(a)}(t)\}
\end{equation*}
where $\hat{E}\{A^{(a)}_k(t)\}$ is the area under the KM curve $\hat{S}^{(a)}_k(\cdot)$ up to time $t$.

Standard errors (SEs) are computed using influence function methodology, with asymptotic normality enabling confidence interval construction and hypothesis testing. The asymptotic expansion of $\hat{E}\{A^{(a)}(t)\}$ is given by
\[
\hat{E}\{A^{(a)}(t)\} - E\{A^{(a)}(t)\}
= \frac{1}{n_a} \sum_{i=1}^{n_a} \xi^{(a)}_i(t) + o_p(1),
\]
where $o_p(1)$ denotes a term that converges to zero in probability as $n_a \to \infty$. The influence function for individual $i$ in arm $a$ is:
\begin{equation*}
\xi_i^{(a)}(t) = \sum_{k=1}^{K+1} \int_0^t \frac{\int_u^t S_k^{(a)}(v)dv}{P(X_{ik}^{(a)} > u)} dM_{ik}^{(a)}(u)
\end{equation*}
where $M_{ik}^{(a)}(t)$ represents the counting process martingale: $M^{(a)}_{ik}(t)
= I\{X^{(a)}_{ik} \le t\}\,\delta^{(a)}_{ik} 
- \int_{0}^{t} I\{X^{(a)}_{ik} \ge u\}\,\lambda^{(a)}_k(u)\,du$, and $\lambda^{(a)}_k(u)$ denotes the hazard function of $T^{(a)}_k$. This framework provides robust variance estimation accounting for censoring and correlation among transition times \citep{claggett2018quantifying,sun2025improve}.

Let $\hat{\sigma}^{(a)}_{A}(t)$ denote the estimated SE of $\hat{E}\{A^{(a)}(t)\}$, defined as $\hat{\sigma}^{(a)}_{A}(t)
= \left[ E\left\{\hat{\xi}^{(a)}_i(t)^2\right\}\big/n_a \right]^{1/2},$ where $\hat{\xi}^{(a)}_i(t)$ is obtained by replacing the unknown quantities in $\xi^{(a)}_i(t)$ with their estimated counterparts (e.g., substituting $\lambda^{(a)}_k(u)$ and $S^{(a)}_k(u)$ by their respective Nelson-Aalen and KM estimators). 

The estimates $\hat R(\tau)$ and $\hat D(\tau)$ are obtained by plugging in the empirical estimates of $E\{A^{(a)}(\tau)\}$. SEs, confidence intervals (CIs), and $p$-values for $\hat R(\tau)$ and $\hat D(\tau)$ are then derived using the $\hat{\sigma}^{(a)}_{A}(t)$ for each arm, with the delta method applied to obtain the SE of $\hat R(\tau)$.

\subsection{Restricted Mean Time in Favor of Treatment (RMT-IF)}
The RMT-IF quantifies treatment benefit as net time spent in more favorable health states over a prespecified follow-up window $[0, \tau]$. For independent processes $Y^{(1)}$ and $Y^{(0)}$ from treatment and placebo arms, we define gross win time as:
\begin{equation*}
W\{Y^{(a)}, Y^{(1-a)}\}(\tau) = \int_0^\tau I\{Y^{(a)}(t) < Y^{(1-a)}(t)\}dt
\end{equation*}

This measures total time during which a patient from arm $a$ maintains a more favorable state than their counterpart. The RMT-IF is then:
\begin{equation*}
\mu(\tau) = E[W\{Y^{(1)}, Y^{(0)}\}(\tau)] - E[W\{Y^{(0)}, Y^{(1)}\}(\tau)]
\end{equation*}
representing net average time advantage conferred by treatment. Positive values indicate treatment benefit through extended time in favorable states.

The RMT-IF admits natural decomposition by disease stage:
\begin{equation*}
\mu(\tau) = \sum_{k=1}^{K+1} \mu_k(\tau)
\end{equation*}
where $\mu_k(\tau)$ quantifies stage-specific treatment effects---the restricted mean time favoring treatment for avoiding state $k$ while remaining free of more severe states. Notably, $\mu_{K+1}(\tau)$ corresponds to the familiar RMST difference. Detailed estimation and inference procedures are provided in \cite{mao2023restricted} and \cite{mao2024dissecting}.

\subsection{Unified Utility Framework}
Both AUC and RMT-IF represent special cases of a general utility-based framework \citep{buhler2025specification}. Defining utility $u_k^{(a)}(t)$ for state $k$ at time $t$ in arm $a$, cumulative utility becomes:
\begin{equation*}
U^{(a)}(\tau) = \sum_{k=1}^{K+1} \int_0^\tau u_k^{(a)}(t)I\{Y^{(a)}(t) = k\}dt
\end{equation*}

The AUC emerges when $u_k^{(a)}(t) = k$ (fixed ordinal scores), while RMT-IF corresponds to $u_k^{(a)}(t) = I\{Y^{(1-a)}(t) < k\}$ (comparative state occupancy). This unifying perspective highlights how different utility specifications capture distinct aspects of treatment benefit.

\subsection{Clinical Application: REWIND Trial Analysis}

We applied both methods to the REWIND trial (NCT00967798) \citep{gerstein2019dulaglutide}, a randomized controlled trial comparing once-weekly dulaglutide 1.5 mg with placebo in 9,901 adults with type 2 diabetes. Our analysis focused on the prespecified subgroup with CKD comprising participants with baseline eGFR $<$ 60 mL/min/1.73 m$^2$ or UACR $>$ 30 mg/g, yielding 4,133 participants (2,040 dulaglutide; 2,093 placebo) with median follow-up of 5.3 years.

We defined five hierarchical renal outcomes based on current regulatory guidance and validation studies:
\begin{enumerate}
\item 40\% eGFR decline from baseline (earliest validated surrogate)
\item 50\% eGFR decline from baseline (intermediate threshold)
\item 57\% eGFR decline from baseline (equivalent to doubling serum creatinine)
\item End-stage renal disease (dialysis, transplantation, or eGFR $<$ 15 mL/min/1.73 m$^2$)
\item All-cause mortality
\end{enumerate}

All analyses followed intention-to-treat principles with $\tau = 6$ years for primary comparisons. We conducted sensitivity analyses systematically excluding less severe eGFR thresholds to assess robustness of treatment effect estimates across endpoint definitions.

\section{Results}\label{sec:results}

\subsection{Baseline Characteristics and Event Distribution}

The CKD risk subgroup showed well-balanced baseline characteristics across treatment arms. Mean age was 67.3 vs.\ 67.1 years, with 45\% vs.\ 46\% female participants. Baseline kidney function indicated moderate CKD risk, with mean eGFR of 70.8 vs.\ 71.7 mL/min/1.73 m$^2$ and median UACR of 78 vs.\ 78 mg/g. Cardiovascular risk factors were common, with 31.8\% vs.\ 31.2\% of participants having a history of cardiovascular events.

Table \ref{tab:first} presents the distribution of first events by type, revealing important patterns obscured by conventional analysis. It is important to note that due to periodic assessment schedules, some participants may first be observed with 50\% or 57\% eGFR decline without prior documentation of smaller declines. In such cases, while these patients have necessarily passed through the 40\% (and for 57\%, also the 50\%) threshold, Table \ref{tab:first} categorizes them by their first observed event only. Thus, a patient whose first observed event is a 50\% decline is counted in that column, not in the 40\% decline column, even though they must have experienced a 40\% decline. Despite this observation pattern, 188 dulaglutide and 180 placebo participants experienced 40\% eGFR decline as their first observed event, with subsequent first observed progression differing markedly between arms.

\begin{table}[htbp]
\centering
\caption{Distribution of First Events by Type in the CKD Risk Subgroup}
\label{tab:first}
\begin{tabular}{lcccccc}
\toprule
Treatment Arm & 40\% Decline & 50\% Decline & 57\% Decline & ESRD & Death & Total Events \\
\midrule
Dulaglutide & 188 & 38 & 43 & 63 & 269 & 601 \\
Placebo & 180 & 57 & 49 & 72 & 311 & 669 \\
\bottomrule
\end{tabular}
\end{table}

Table \ref{tab:worst} reveals the complete disease trajectory by showing each participant's worst observed state. Relative to Table \ref{tab:first}, 121 dulaglutide versus 114 placebo participants experienced only 40\% decline without further progression, while 67 and 66 participants respectively progressed beyond their first event---information entirely missed by time-to-first analysis.

\begin{table}[htbp]
\centering
\caption{Distribution of Worst Observed Renal State Per Participant in the CKD Risk Subgroup}
\label{tab:worst}
\begin{tabular}{lccccccc}
\toprule
Treatment Arm & Censored & 40\% Only & 50\% Max & 57\% Max & ESRD & Death & Total \\
\midrule
Dulaglutide & 1438 & 121 & 43 & 50 & 56 & 332 & 2040 \\
Placebo & 1422 & 114 & 46 & 73 & 68 & 370 & 2093 \\
\bottomrule
\end{tabular}
\end{table}

\subsection{Conventional Time-to-First Analysis}

\begin{figure}[htbp]
  \centering
  \includegraphics[width=0.7\textwidth]{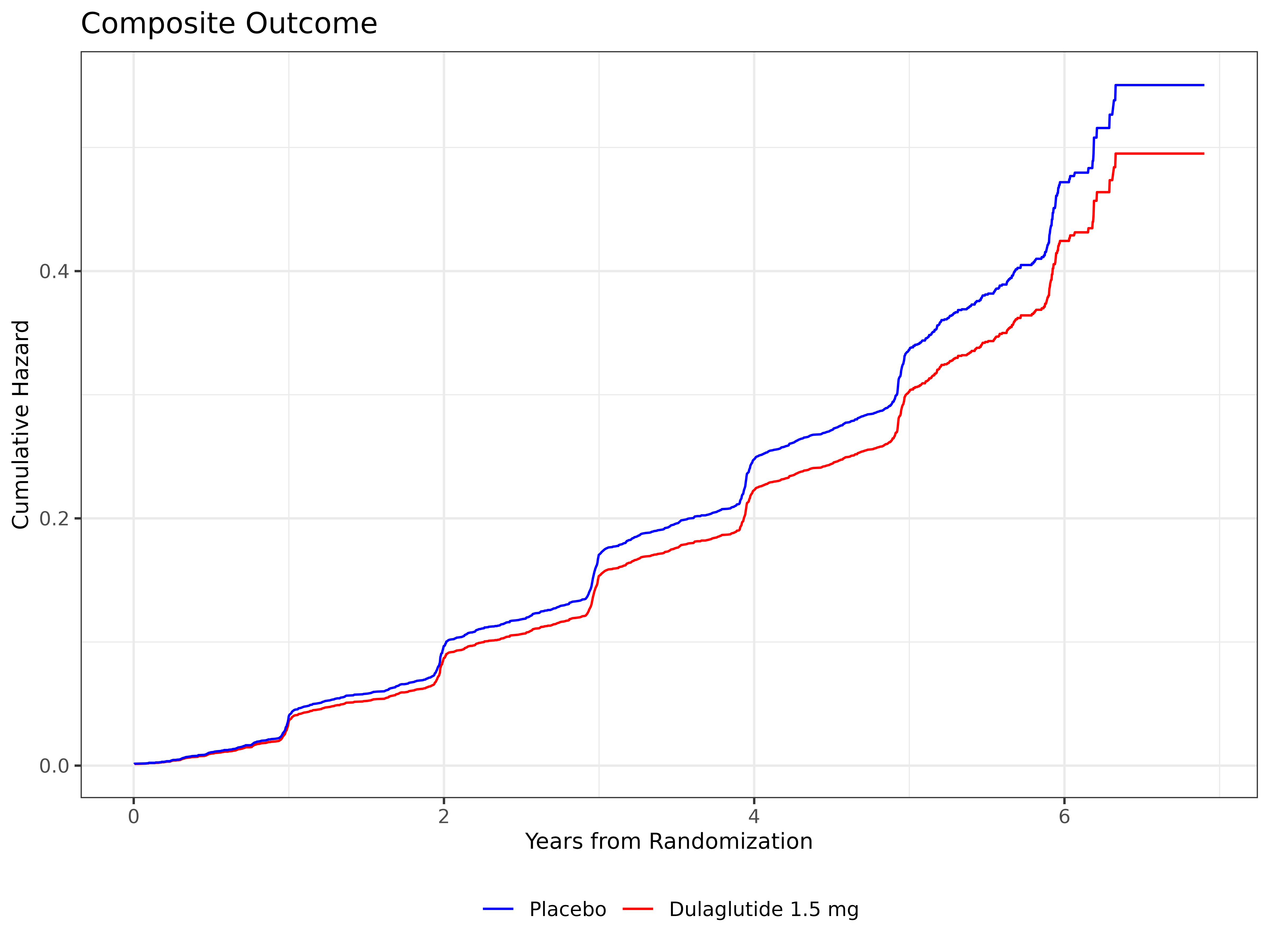}
  \caption{Cumulative hazard curves for the first occurrence of composite renal outcomes, including death, ESRD, and eGFR declines of $57\%$, $50\%$, and $40\%$ from baseline, by treatment arm in the CKD risk subgroup of the REWIND trial.}
  \label{fig:composite_outcome}
\end{figure}

The composite endpoint including all eGFR decline (40\%, 50\%, or 57\%), ESRD, or death yielded a hazard ratio of 0.899 (95\% CI: 0.806--1.004; $p = 0.06$), narrowly missing statistical significance at $\alpha = 0.05$. The cumulative hazard curves for both treatment arms are shown in Figure~\ref{fig:composite_outcome}.

\subsection{Area Under the Curve Analysis}

\begin{figure}[htbp]
  \centering
  \includegraphics[width=0.7\textwidth]{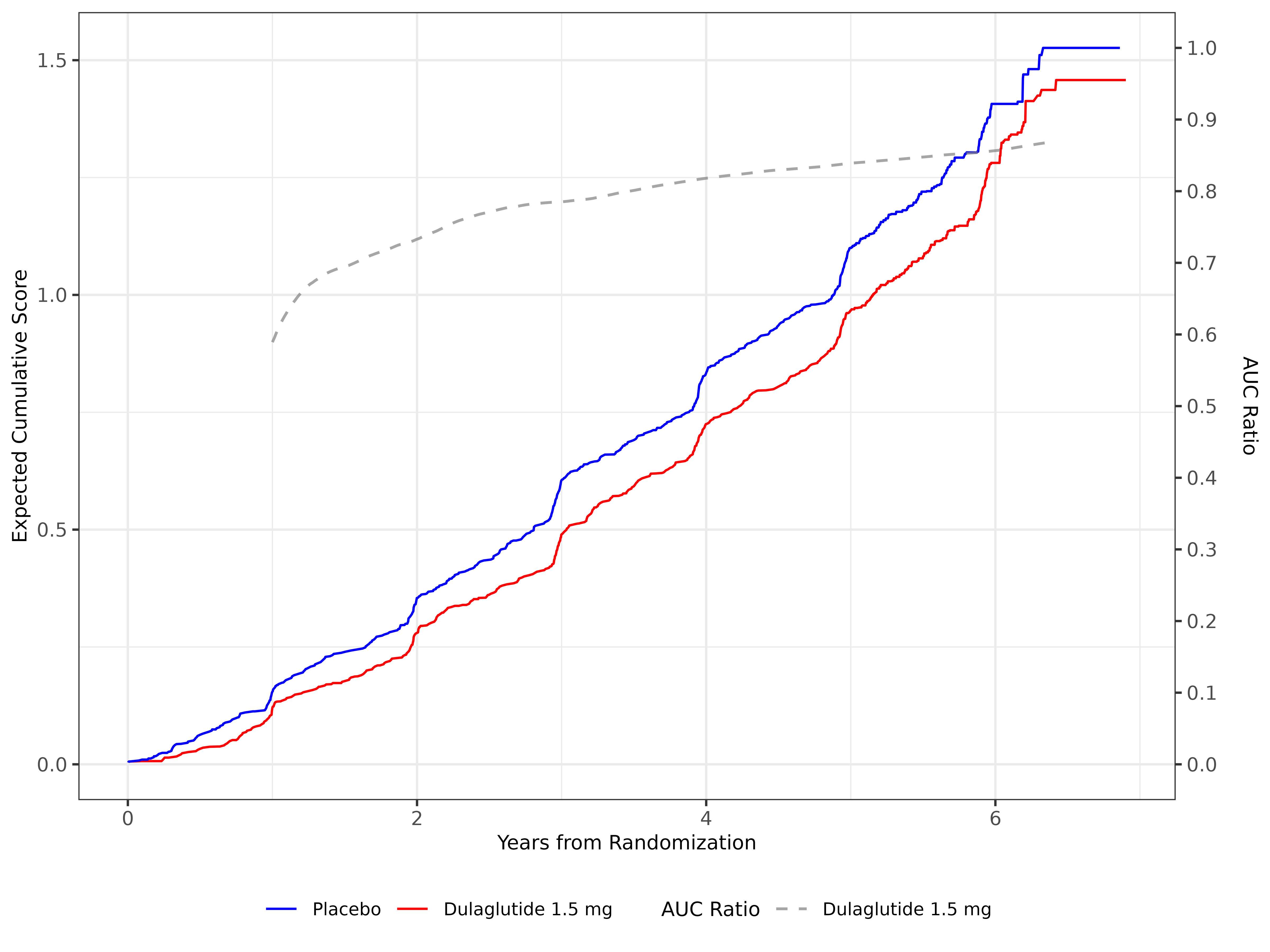}
  \caption{Mean cumulative disease score curves by treatment arm in the CKD risk subgroup of the REWIND trial. Higher scores indicate greater disease burden. The gray dotted line shows the AUC ratio over time.}
  \label{fig:auc}
\end{figure}

Figure \ref{fig:auc} displays the evolution of mean cumulative disease scores over time. The curves diverged early and maintained consistent separation, with dulaglutide demonstrating lower cumulative burden throughout follow-up. At six years, the AUC ratio was 0.857 (95\% CI: 0.760--0.966; $p = 0.012$), representing a statistically significant 14.3\% reduction in cumulative disease burden.

The absolute AUC difference was $-0.516$ event-years (SE $= 0.204$; 95\% CI: $-0.916$ to $-0.115$; $p = 0.012$), interpretable as prevention of approximately 6 event-months of disease burden per patient over 6 years. Decomposition by component revealed contributions from each transition: extended time free of 40\% decline (0.112 years), 50\% decline (0.123 years), 57\% decline (0.106 years), ESRD (0.095 years), and extended survival (0.080 years).

\subsection{Restricted Mean Time in Favor of Treatment Analysis}

Table \ref{tab:rmtif} presents the RMT-IF decomposition, revealing stage-specific treatment benefits. The overall RMT-IF of 0.116 years (95\% CI: 0.021--0.211; $p = 0.018$) indicates dulaglutide-treated patients in a CKD risk subgroup spent an average of 42 additional days in more favorable health states compared to placebo, over 6 years.

\begin{table}[htbp]
\centering
\setlength{\tabcolsep}{14pt} 
\caption{Restricted Mean Time Difference (in Years, $\tau=6$) in Favor of Dulaglutide in the CKD Risk Subgroup}
\label{tab:rmtif}
\begin{tabular}{lccc}
\toprule
Outcome Avoided & Estimate & SE & $P$-value \\
\midrule
40\% eGFR Decline & $-0.004$ & 0.018 & 0.799 \\
50\% eGFR Decline & 0.016 & 0.009 & 0.070 \\
57\% eGFR Decline & 0.009 & 0.011 & 0.445 \\
ESRD & 0.016 & 0.016 & 0.331 \\
Death (RMST) & 0.080 & 0.039 & 0.039 \\
\midrule
\textbf{Overall RMT-IF} & \textbf{0.116} & \textbf{0.049} & \textbf{0.018} \\
\bottomrule
\end{tabular}
\end{table}

The largest component was extended survival (0.080 years, representing 69\% of total benefit), followed by time free of ESRD among survivors (0.016 years). The 40\% decline component showed a small negative value ($-0.004$ years), though not statistically significant, suggesting comparable early progression rates between arms with divergence occurring at more severe stages.

\subsection{Comparative Interpretation of Results}

Both novel approaches detected statistically significant treatment benefits where conventional analysis did not. The 14.3\% reduction in cumulative disease burden as reflected in the total event-time that would have been lost due to CKD disease progression (AUC) and 42 additional days in favorable health states (RMT-IF) provide complementary perspectives on treatment effect magnitude. These measures offer more intuitive clinical interpretation than hazard ratios, directly quantifying the patient experience of disease progression.

The decomposition capabilities of both methods revealed that survival extension contributed most substantially to overall treatment benefit (69\% of RMT-IF effect), followed by delays in progression to ESRD. This granular insight into stage-specific effects would be impossible with conventional composite endpoint analysis.

\subsection{Sensitivity Analysis: Impact of Endpoint Definition}

We systematically evaluated treatment effect estimates while progressively excluding less severe eGFR thresholds (Table \ref{tab:sensitivity}). This analysis revealed striking differences in stability between analytical approaches. For the AUC-based analysis, the ordinal score assigned to each state follows the same principle: death is always assigned the value $K+1$, whereas the value of $K$ depends on the number of events included in the composite endpoint.

\begin{table}[htbp]
\centering
\setlength{\tabcolsep}{7pt}   
\renewcommand{\arraystretch}{1.3} 
\caption{Treatment Effect Estimates Across Different Endpoint Definitions}
\label{tab:sensitivity}
\small
\begin{tabular}{lcccccc}
\toprule
Endpoints Included & Cox HR (95\% CI) & $P$ & AUC Ratio (95\% CI) & $P$ & RMT-IF (years) & $P$ \\
\midrule
All five endpoints & 0.899 (0.806--1.004) & 0.059 & 0.857 (0.760--0.966) & 0.012 & 0.116 & 0.018 \\
Exclude 40\% decline & 0.864 (0.765--0.976) & 0.019 & 0.848 (0.745--0.965) & 0.013 & 0.120 & 0.008 \\
Exclude 40\%, 50\% declines & 0.861 (0.758--0.978) & 0.022 & 0.851 (0.744--0.974) & 0.019 & 0.104 & 0.017 \\
ESRD + death only & 0.883 (0.770--1.013) & 0.076 & 0.851 (0.739--0.980) & 0.025 & 0.096 & 0.023 \\
\bottomrule
\end{tabular}
\end{table}

The AUC ratio remained stable (0.851--0.857) regardless of included thresholds, with consistent statistical significance maintained throughout. This stability reflects the method's weighting scheme that naturally emphasizes more severe events. In contrast, Cox hazard ratios varied substantially (0.861--0.899), with statistical significance fluctuating based on endpoint composition. Removing the 40\% threshold paradoxically improved statistical significance in time-to-first analysis, as dulaglutide observed slightly higher rates of early 40\% decline as first events.

\section{Discussion}\label{sec:discussion}
Our analysis of the REWIND trial demonstrated that incorporating disease severity and state duration provided alternative characterization of treatment effects in CKD trials. In this specific example, where conventional time-to-first analysis found marginal effects ($p = 0.06$), both AUC and RMT-IF approaches revealed statistically significant treatment benefits. The choice of analytical approach depends on the underlying clinical question of interest. When the goal is to understand treatment effects on delaying the first event, conventional time-to-first analysis remains appropriate. However, when the objective is to capture the totality of disease burden in progressive conditions, methods that integrate both event severity and duration may provide additional insights not captured by traditional approaches.

In the REWIND study context, the 14.3\% reduction in cumulative disease burden represents a substantial clinical benefit extending beyond simple event delay. For healthcare systems and patients, this translates to reduced resource utilization across the spectrum of CKD care, fewer progressions to dialysis-requiring ESRD, and extended productive life years. The 42 additional days in favorable health states, while modest in absolute terms, represents meaningful quality-of-life preservation when considered across populations and over treatment lifetimes.

Within the REWIND sensitivity analysis, the stability of AUC estimates across different endpoint definitions proved particularly noteworthy. This robustness addresses a persistent challenge in CKD trials: the selection of appropriate eGFR decline thresholds. Our findings in this trial suggest that when more severe events are present, the inclusion of validated earlier thresholds (40\%, 50\%) enhances statistical power without substantially altering AUC effect estimates---a property not demonstrated by the conventional Cox models in this analysis. However, this stability may vary across different trial populations and treatment effects.

The AUC approach offers several key advantages for CKD trials. First, its explicit severity weighting aligns with clinical intuition---ESRD carries greater weight than early eGFR decline, reflecting true clinical impact. Second, the measure provides absolute quantification of burden in intuitive units (event-years), facilitating communication with patients, clinicians, and policymakers. Third, the method's stability across endpoint specifications reduces concerns about post-hoc endpoint selection while maintaining sensitivity to detect treatment effects.

The RMT-IF complements AUC by providing a pairwise comparison framework familiar from win ratio analyses while incorporating time dimensions. Its decomposition into stage-specific components revealed where in the disease trajectory treatments exert greatest benefit---critical information for understanding mechanisms and optimizing therapeutic strategies. The finding that survival extension dominated overall benefit (69\% of effect) suggests dulaglutide's effects extend beyond pure renal protection to broader mortality reduction.

Both methods share the advantage of not requiring proportional hazards assumptions, which are frequently violated in CKD trials. This model-free nature ensures valid inference even with non-proportional treatment effects. While traditional analyses can estimate hazard ratios for individual components such as mortality, the proposed approach complements these analyses by jointly integrating multiple clinically ordered events into a single estimand of cumulative disease burden. At the same time, both AUC and RMT-IF naturally admit decomposition into component- or stage-specific contributions, allowing investigators to recover and interpret individual outcome effects within a unified framework. This enables treatment benefits to be assessed holistically while retaining insight into differential effects across endpoints and over time that may not be apparent from traditional separate component-wise analyses.

Our findings have important implications for designing future CKD trials. The enhanced statistical power of AUC and RMT-IF analyses could enable smaller, shorter trials while maintaining ability to detect clinically meaningful effects. The stability of estimates across endpoint definitions provides flexibility in trial design, allowing inclusion of earlier surrogate endpoints without compromising interpretability. Regulatory agencies increasingly recognize the limitations of time-to-first analyses for chronic diseases. Our results provide empirical support for incorporating AUC or RMT-IF as additional evidence in CKD outcome trials, potentially accelerating development of kidney-protective therapies.

While demonstrated here for CKD, these analytical approaches apply broadly to chronic progressive diseases. Conditions such as heart failure, chronic obstructive pulmonary disease, and neurodegenerative disorders share similar characteristics: predictable progression through defined stages, clear severity hierarchies, and treatment goals extending beyond first-event prevention. The methods could also enhance evaluation of preventive interventions where cumulative benefit accrues over extended periods. In oncology, where progression-free survival inadequately captures treatment benefit for therapies affecting post-progression outcomes, AUC approaches could integrate multiple progression events and survival into comprehensive efficacy measures. Similarly, in diabetes complications trials, these methods could simultaneously evaluate microvascular and macrovascular outcomes while respecting their relative clinical importance.

Several limitations merit consideration. First, ordinal scoring in AUC analysis requires clinical judgment and consensus. While our scoring (1-5 for increasing severity) has face validity, alternative schemes might yield different results. In certain clinical contexts, a more clinically relevant severity rating---defined using medical judgment and aligned with specific disease or treatment considerations---may be preferable.  Sensitivity analyses with varied scoring systems would strengthen conclusions, though our stability analyses suggest robust findings. Second, both methods require complete follow-up or appropriate censoring mechanisms. Informative censoring, where dropout relates to disease progression, could bias estimates. However, the REWIND trial's high retention rate ($>$95\%) minimizes this concern. Third, the clinical interpretation of effect magnitudes requires contextualization. While 42 additional days in favorable states appears modest, this represents population averages including many non-progressors. Among those experiencing events, individual benefits may be substantially larger. Fourth, the methods assume irreversible disease progression, appropriate for CKD but potentially limiting for conditions with improvement potential. Extensions accommodating bidirectional transitions exist but increase complexity. Finally, implementation requires statistical expertise beyond standard survival analysis. However, available software packages increasingly support these analyses, and the clinical and analytical advantages might justify the additional investment.

Several avenues warrant further investigation. Incorporating baseline covariates could enhance precision while maintaining marginal interpretability---recent methodological advances in nonparametric covariate adjustment offer promising approaches \citep{sun2025improve,ye2022inference,ye2023toward,ye2024covariate,FDA2023}. Development of sample size calculation methods specific to AUC/RMT-IF endpoints would facilitate trial planning. Investigation of optimal ordinal scoring systems, potentially through utility elicitation studies with patients and clinicians, could standardize approaches across trials. Extensions to competing risks settings would broaden applicability, particularly for interventions with differential effects on renal versus cardiovascular outcomes.

The management of chronic kidney disease demands analytical approaches that reflect its progressive nature and cumulative impact on patients' lives. Our application of area under the curve and restricted mean time in favor of treatment methods to the REWIND trial demonstrates their ability to capture treatment benefits invisible to conventional time-to-first analyses. By integrating event severity with duration in each disease state, these approaches provide comprehensive, clinically interpretable measures of therapeutic efficacy.

The transition from time-to-first-event to cumulative burden assessment represents more than methodological refinement---it reflects fundamental recognition that chronic diseases impact patients through their entire trajectory, not merely through initial manifestations. By quantifying this complete journey, we move closer to treatments that meaningfully alter disease course and improve patients' lives throughout their illness experience.


\bibliography{maintext}  

\clearpage

\end{document}